\documentclass[preprint,prb,aps,floats,amssymb,showpacs,draft,%
floatfix,superscriptaddress]{revtex4} %,nofootinbib
\usepackage{psfig}

\begin{document}

\title{
Universal dependence on disorder of
2D randomly diluted and random-bond $\pm J$ Ising models
}
\author{Martin Hasenbusch} 
\affiliation{ 
Institut f\"ur Theoretische Physik, Universit\"at Leipzig, 
Postfach 100 920, D-04009 Leipzig, Germany.
} 
\author{Francesco Parisen Toldin} 
\affiliation{ 
Max-Planck-Institut f\"ur Metallforschung,
Heisenbergstrasse 3, D-70569 Stuttgart, Germany\\
and Institut f\"ur Theoretische und Angewandte Physik,
Universit\"at Stuttgart,
Pfaffenwaldring 57, D-70569 Stuttgart, Germany
} 
\author{Andrea Pelissetto} 
\affiliation{Dipartimento di Fisica
  dell'Universit\`a di Roma ``La Sapienza" and INFN, Roma, Italy.}
\author{Ettore Vicari} 
\affiliation{ 
Dipartimento di Fisica dell'Universit\`a di Pisa and INFN, Pisa, Italy.  } 

\date{\today}

\begin{abstract}
  We consider the two-dimensional randomly site diluted Ising model and the
  random-bond $\pm J$ Ising model (also called Edwards-Anderson model), 
  and study their critical behavior at the
  paramagnetic-ferromagnetic transition.  The critical behavior of
  thermodynamic quantities can be derived from a set of renormalization-group
  equations, in which disorder is a marginally irrelevant perturbation at the
  two-dimensional Ising fixed point.  We discuss their solutions, focusing in
  particular on the universality of the logarithmic corrections arising from
  the presence of disorder. Then, we present a finite-size scaling analysis of
  high-statistics Monte Carlo simulations. The numerical results confirm the
  renormalization-group predictions, and in particular the universality of the
  logarithmic corrections to the Ising behavior due to quenched dilution.
\end{abstract}


%% 75.10.Nr Spin-glass and other random models
%% 75.40.Cx Static properties (order parameter, static susceptibility, 
%%          heat capacities, critical exponents, etc.)
%% 75.40.Mg Numerical simulation studies
%% 64.60.Fr Equilibrium properties near critical points, critical exponents

\maketitle

% ========================= BODY =========================
%\narrowtext

\section{Introduction and summary}

Random Ising systems represent an interesting theoretical laboratory in which
one can study general features of disordered systems.  Among them, the
two-dimensional (2D) random-site and random-bond Ising models have attracted
much interest. In particular, the effects of quenched disorder on the critical
behavior at the paramagnetic-ferromagnetic transitions, which are observed for
sufficiently small disorder, have been much investigated and debated, see
Refs.~\onlinecite{DD-81,Shalaev-84,Cardy-86,Shankar-87,Ludwig-87,%%
  LC-87,Mayer-89,Ludwig-90,%%
  Ziegler-90,WSDA-90,Heuer-92,Shalaev-94,KP-94,Kuhn-94,QS-94,TS-94,TS-94-2,%%
  MS-95,DPP-95,AQS-96,JS-96,CHMP-97,BFMMPR-97,AQS-97,SSLI-97,%%
  RAJ-98,SSV-98,AQS-99,MK-99,LSZ-01,Nobre-01,SV-01,MC-02,%%
  COPS-04,Queiroz-06,LQ-06,PHP-06,MP-07,KJJ-07}.

Renormalization-group (RG) and conformal field theory
\cite{Shalaev-84,Shankar-87,Ludwig-90,Shalaev-94} predict the marginal
irrelevance of random dilution at the paramagnetic-ferromagnetic transition.
Therefore, the asymptotic behavior is controlled by the standard Ising fixed
point, characterized by the critical exponents $\nu=1$ and $\eta=1/4$;
disorder gives only rise to logarithmic corrections.  The marginality of
quenched disorder coupled to the energy density, as it is the case for random
dilution, is already suggested by the Harris criterion,\cite{Harris-74} which
states that the relevance or irrelevance of quenched dilution depends on the
sign of the specific-heat exponent of the pure system; in the case of the 2D
Ising model, the specific heat diverges only logarithmically at the
transition, i.e. $\alpha=0^+$.  The marginal irrelevance of disorder has also
been supported by numerical studies of lattice models; see, e.g.,
Refs.~\onlinecite{WSDA-90,BFMMPR-97,AQS-97,RAJ-98,MK-99,SV-01,MC-02,COPS-04,%%
  LQ-06,PHP-06,MP-07} (see, however, Refs.~\onlinecite{KP-94,Kuhn-94,LSZ-01}
for different scenarios).  We recall that in three dimensions random dilution
is a relevant perturbation of the pure Ising fixed point, leading to
a new three-dimensional randomly diluted Ising (RDI) universality class, which
is characterized by different critical exponents; see, e.g.,
Refs.~\onlinecite{PV-02,HPPV-07}.

In this paper we return to the issue of the critical behavior of 2D randomly
diluted Ising systems.  By using the RG results reported in
Refs.~\onlinecite{Cardy-86,Ludwig-87,LC-87,MK-99}, we show that their critical
behavior can be derived from the RG equations
\begin{eqnarray}
&& {d u_I\over d l} = 2 u_I + d_I u_t^2 ,  \nonumber \\
&& {d u_t\over d l} = u_t - {1\over 2} g u_t ,  \nonumber \\
&& {d u_h\over d l} = {15\over 8} u_h , \nonumber \\
&& {d g\over d l} = -g^2 + {1\over 2} g^3, \label{gop}
\end{eqnarray}
where $l$ is the flow parameter (the logarithm of a length scale),
$u_I$, $u_t$, and $u_h$ are the scaling fields associated with the
leading operators of the three different conformal families of the 2D Ising
model, i.e. the identity, energy, and spin families, and $g$ is the marginally
irrelevant scaling field associated with disorder.  Higher-order terms in
Eqs.~(\ref{gop}) are not necessary, because they can be reabsorbed by
appropriate analytic redefinitions of the scaling fields. The appearance of
the term $d_I u_t^2$ in the first equation, where $d_I$ is a nonuniversal
constant, is due to the resonance of the identity operator with the thermal
operator, which already occurs in the pure Ising model.\cite{Wegner-76} It is
interesting to note that randomness couples only to the thermal scaling field
$u_t$.  It would be interesting to understand if these conclusions also apply
to the irrelevant operators, i.e., if the only operators that couple disorder
are those that belong to the conformal family of the energy.

The analysis of the RG equations shows that random dilution gives rise to
logarithmic corrections which are universal after an appropriate normalization
of the scaling field associated with disorder.  Additional scaling corrections
due to the irrelevant operators are suppressed by power laws as in standard
continuous transitions. For these reasons, we prefer to distinguish the
randomly dilute Ising (RDI) critical behavior characterized by the RG
equations (\ref{gop}) from the standard 2D Ising universality class of pure
systems, although they share the same 2D Ising fixed point.

The RG equations (\ref{gop}) allow us to determine the scaling behavior of any
thermodynamic quantity. Denoting with $t$, $h$, $p$, and $L$ the reduced
temperature, the magnetic external field, the disorder parameter, and the
lattice size, respectively, the free energy satisfies the scaling equation
\begin{equation}
F(t,h,p,L) = e^{-2l} u_I(l) + e^{-2l} f(u_t(l),u_h(l),g(l),e^l L)
\label{scaleq}
\end{equation}
(we consider models defined on square $L\times L$ lattices with periodic
boundary conditions), where $u_I(l)$, $u_t(l)$, $u_h(l)$, and $g(l)$ are the
solutions of the RG equations.  From Eq.~(\ref{scaleq}) one can derive the
scaling behavior of the relevant thermodynamic quantities and determine the
logarithmic corrections due to the quenched disorder.  At the critical point
$t=h=0$, we obtain the asymptotic behaviors\cite{MK-99}
\begin{equation}
C_h \sim \ln \ln L 
\label{specheat}
\end{equation}
for the specific heat, and
\begin{eqnarray}
&&\chi = c L^{7/4} f_\chi(g(\ln L)) = 
c L^{7/4} \left[1 + O\left({1\over \ln L}\right)\right],\label{magchi}\\
&& R = R^* f_R(g(\ln L)) = R^*  +  O\left({1\over \ln L}\right),\label{rginv}\\
&&{d R\over d t} = c L g(\ln L)^{1/2} f_{dR}(g(\ln L))
= c {L \over \sqrt{\ln L}} \left[1 + O\left({1\over \ln L}\right)\right]
\label{rginvder}
\end{eqnarray}
for the magnetic susceptibility $\chi$, any RG invariant quantity $R$, such as
the quartic Binder cumulant $U_4$ and the ratio $R_\xi\equiv \xi/L$, and its
derivative with respect to the temperature.  Here $R^*$ indicates the Ising
fixed-point value and $g(\ln L)$ is the solution of Eq.~(\ref{gop}) with $l =
\ln L$. For $L\to \infty$, $g(\ln L)$ behaves as
\begin{equation}
g(\ln L) \sim {1\over \ln L/L_0} \left[
1   +{\ln \ln {L/L_0} \over 2 \ln L/L_0}  + \cdots \right],
\label{gsc}
\end{equation}
where $L_0$ is a length scale.  The functions $f_\chi(x)$, $f_R(x)$, and
$f_{dR}(x)$ are normalized such that $f_\#(0)=1$ and are universal once the
scaling field $g(\ln L)$ is appropriately normalized.  In the above-reported
equations we have neglected scaling corrections which are suppressed by power
laws. They are due to the analytic dependence of the scaling fields on the
Hamiltonian parameters, to the background (i.e., the contribution of the
identity operator in the RG language), and to the irrelevant
operators.~\cite{CCCPV-00,CHPV-02} In particular, we expect scaling
corrections associated with the leading irrelevant operator appearing in the
pure Ising model (the corresponding exponent is $\omega=2$).

Moreover, in this paper we compare the theoretical predictions with a
finite-size scaling (FSS) analysis of numerical Monte Carlo (MC) results for
the randomly site-diluted Ising model and for the random-bond $\pm J$ Ising
model, also known as Edwards-Anderson model.  Our main results can be
summarized as follows.  Our FSS analyses provide a robust evidence that the
paramagnetic-ferromagnetic transitions in these models present the same RDI
critical behavior.  Note that this implies that frustration in the random-bond
$\pm J$ Ising model is irrelevant at the paramagnetic-ferromagnetic transition
line. The FSS behaviors are in agreement with the predictions of the RG
equations (\ref{gop}).  The asymptotic critical behavior appears to be
controlled by the Ising fixed point.  The logarithmic corrections and their
universal behavior are consistent with the theoretical results obtained from
Eqs.~(\ref{gop}), cf. Eqs.~(\ref{scaleq}), (\ref{specheat}), (\ref{magchi}),
(\ref{rginvder}), (\ref{gsc}).

The paper is organized as follows. In Sec.~\ref{sec2} we define the
randomly site-diluted model and the $\pm J$ Ising model, we briefly
discuss their phase diagrams, and define the quantities that we consider in
the paper.  In Sec.~\ref{RGFSS} we discuss the RG flow at a 2D Ising fixed
point in the presence of a marginally irrelevant operator, and the
implications for the infinite-volume and finite-size critical behavior. In
particular, we focus on the universal features of the logarithmic corrections
due to disorder.  Finally, in Sec.~\ref{secMC} we present our FSS analysis of
high-statistics MC results for the randomly site-diluted and random-bond $\pm
J$ Ising models.

\section{Randomly site-diluted and random-bond $\pm J$ Ising models}
\label{sec2}

\subsection{The models and their phase diagrams}
\label{models}

The randomly site-diluted Ising model (RSIM) is defined by the Hamiltonian
\begin{equation}
{\cal H}_{s} = - \sum_{<xy>}  \rho_x \,\rho_y \; \sigma_x \sigma_y,
\label{Hs}
\end{equation}
where the sum is extended over all pairs of nearest-neighbor sites of a
square lattice, $\sigma_x=\pm 1$ are Ising spin variables, and $\rho_x$ are
uncorrelated quenched random variables, which are equal to one with
probability $p$ (the spin concentration) and zero with probability $1-p$ (the
impurity concentration).  The RSIM is expected to undergo a continuous
transition for any $p>p_{\rm perc}$, where\cite{NZ-00} $p_{\rm
  perc}=0.59374621(13)$ corresponds to the site-percolation point of the
impurities; moreover, $T_c\to 0$ for $p\to p_{\rm perc}$, see, e.g.,
Ref.~\onlinecite{Cardy-book}. For $p\le p_{\rm perc}$ the ferromagnetic phase
is absent. Thus, the paramagnetic-ferromagnetic transition line starts from
the pure Ising point $X_{\rm Is}=(p=1,T=T_{\rm Is})$, where $T_{\rm
  Is}=2/\ln(1+\sqrt{2})=2.26919...$ is the critical temperature of the 2D
Ising model, and ends at $X_{\rm perc}=(p=p_{\rm perc},T=0)$.  Along this line
the critical behavior is expected to be universal, i.e.~independent of
dilution, and to be characterized by the
RG equations (\ref{gop}).  As we shall see, this is supported by
the analysis of our MC results.

\begin{figure*}[tb]
\centerline{\psfig{width=9truecm,angle=0,file=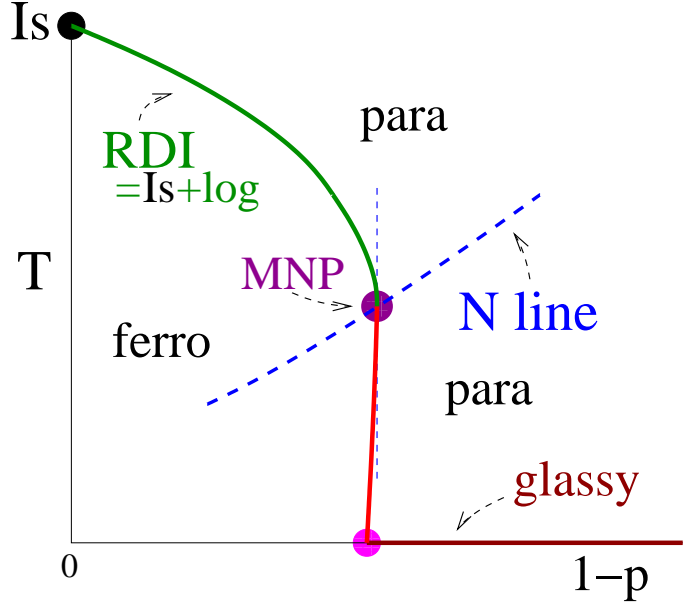}}
\caption{ Phase diagram of the square-lattice random-bond $\pm J$ Ising
  (Edwards-Anderson) model in the $T$-$p$ plane. }
\label{phdia}
\end{figure*}

The random-bond $\pm J$ Ising model, also known as Edwards-Anderson
model,\cite{EA-75} is defined by the lattice Hamiltonian
\begin{equation}
{\cal H}_b = - \sum_{\langle xy \rangle} J_{xy} \sigma_x \sigma_y,
\label{lH}
\end{equation}
where $\sigma_x=\pm 1$, the sum is over all pairs of nearest-neighbor sites of a
square lattice, and the exchange interactions $J_{xy}$ are uncorrelated
quenched random variables, taking values $\pm J$ with probability distribution
\begin{equation}
P(J_{xy}) = p \delta(J_{xy} - J) + (1-p) \delta(J_{xy} + J). 
\label{probdis}
\end{equation}
In the following we set $J=1$ without loss of generality.  For $p=1$ we
recover the standard Ising model, while for $p=1/2$ we obtain the bimodal
Ising spin-glass model.  The $\pm J$ Ising model is a simplified model
\cite{EA-75} for disordered spin systems showing glassy behavior in some
region of their phase diagram.  Its phase diagram in two dimensions 
is sketched in Fig.~\ref{phdia} (it is symmetric for $p\rightarrow
1-p$).  For sufficiently small values of the probability $p_a\equiv 1-p$ of
the antiferromagnetic bonds, the model presents a paramagnetic phase and a
ferromagnetic phase.  The paramagnetic-ferromagnetic transition line starts
from the Ising point $X_{\rm Is}=(p=1,T=T_{\rm Is})$ and extends up to the
multicritical Nishimori point (MNP) $X^*=(p^*,T^*)$, located along the
so-called Nishimori line ($N$ line) defined by
$2p-1={\rm Tanh}(1/T)$,~\cite{Nishimori-81,GHDB-85,LH-88,LH-89} 
with~\cite{HPPV-08}
$p^*=0.89081(7)$ and $T^*=0.9528(4)$.  The critical behavior is 
expected to be in the same universality class as that of the transition in 
the RSIM.  As we shall see, our FSS analysis 
strongly supports this scenario.  A detailed discussion of the phase
diagram can be found in Ref.~\onlinecite{HPPV-08} and references therein.

\subsection{Observables}
\label{observables}

In our FSS analyses we consider models defined on 
a square $L\times L$ lattice with periodic boundary conditions.
The two-point correlation function is defined as
\begin{equation}
G(x) \equiv [ \langle \sigma_0 \,\sigma_x \rangle ],
\label{twof}
\end{equation}
where the angular and square brackets indicate the
thermal average and the quenched average over disorder, i.e. over $\rho_x$ in
the case of RSIM and over $J_{xy}$ in the case of the $\pm J$ Ising model.  We
define the magnetic susceptibility $\chi\equiv \sum_x G(x)$ and the
correlation length $\xi$,
\begin{equation}
\xi^2 \equiv {\widetilde{G}(0) - \widetilde{G}(q_{\rm min}) \over 
          \hat{q}_{\rm min}^2 \widetilde{G}(q_{\rm min}) },
\end{equation}
where $q_{\rm min} \equiv (2\pi/L,0)$, $\hat{q} \equiv 2 \sin q/2$, and
$\widetilde{G}(q)$ is the Fourier transform of $G(x)$.  We also consider
quantities that are invariant under RG transformations in the critical limit.
Beside the ratio
\begin{equation}
R_\xi \equiv \xi/L,
\label{rxi}
\end{equation}
we consider the quartic cumulants $U_4$ and $U_{22}$ defined by
\begin{eqnarray}
U_{4}  \equiv { [ \mu_4 ]\over [\mu_2]^{2}}, 
\qquad U_{22} \equiv  {[ \mu_2^2 ]-[\mu_2]^2 \over [\mu_2]^2},
\label{cumulants}
\end{eqnarray}
where
\begin{eqnarray}
\mu_{k} \equiv \langle \; ( \sum_x \sigma_x\; )^k \rangle \; .
\end{eqnarray}
The above RG invariant quantities $R_\xi$, $U_4$, and $U_{22}$ are also called
phenomenological couplings. 
In the critical ($T=T_c$) 2D pure Ising model, 
they converge for large $L$ to the universal values\cite{SS-00}
\begin{eqnarray}
&&R_\xi^*= R_{\rm Is}=0.9050488292(4),\label{rxii}\\
&&U_4^*= U_{\rm Is} = 1.167923(5),\label{u4i}\\
&&U_{22}^*=0.\label{u22i}
\end{eqnarray}
Finally, we consider the derivatives
\begin{equation}
R'_\xi\equiv {d R_\xi\over d\beta},\qquad
U'_4\equiv {d U_4\over d\beta},
\label{derivatives}
\end{equation}
which can be computed by measuring appropriate expectation values at fixed
$\beta$ and $p$.

\section{Renormalization-group flow and finite-size scaling}
\label{RGFSS}

\subsection{Ising RG flow in the presence of  a marginally 
irrelevant scaling field associated with  disorder}
\label{rgeqrdi}

In this section we discuss the RG flow close to the 2D Ising fixed point in the
presence of a marginally  irrelevant scaling field associated with  disorder.

Let us consider a system with a marginal scaling field $\hat{u}_0
\equiv\hat{g}$ and with a set of scaling fields $\hat{u}_k$, $k\ge 1$, with RG
dimensions $y_k\not=0$. Close to the fixed point $\hat{g} = \hat{u}_1 = \ldots
= 0$, the RG equations have the form
\begin{eqnarray}
&&{d\hat{g}\over dl} = \sum_{0 \le i\le j} b_{0,ij} \hat{u}_i \hat{u}_j + 
        \sum_{0 \le i\le j\le m } b_{0,ijm} \hat{u}_i \hat{u}_j \hat{u}_m +
       \ldots 
\nonumber \\
&&{d\hat{u}_k\over dl} = y_k u_k + 
  \sum_{0\le i\le j} b_{k,ij} \hat{u}_i \hat{u}_j + 
        \sum_{0\le i\le j\le m } b_{k,ijm} \hat{u}_i \hat{u}_j \hat{u}_m +
        \ldots 
\label{RGeq-ini}
\end{eqnarray}
If there are no degeneracies ($y_k \not= y_h$ for all $k\not=
h$) and no resonancies (i.e., 
there is no combination with integer coefficients of
the RG dimensions that vanishes), one can redefine the scaling fields in such a
way to simplify the RG equations. We define
\begin{eqnarray}
&&g =\hat{g} + \sum_{0 \le i\le j} c_{0,ij} \hat{u}_i \hat{u}_j +
        \sum_{0 \le i\le j\le m } c_{0,ijm} \hat{u}_i \hat{u}_j \hat{u}_m +
       \ldots
\\
&&u_k = \hat{u}_k + \sum_{0 \le i\le j} c_{k,ij} \hat{u}_i \hat{u}_j +
        \sum_{0 \le i\le j\le m } c_{k,ijm} \hat{u}_i \hat{u}_j \hat{u}_m +
       \ldots 
\end{eqnarray}
With a proper choice of the coefficients $c_{k,ij}$, $c_{k,ijm}$, $\ldots$,
we can simplify the RG equations, obtaining the simple canonical form:
\begin{eqnarray}
&&{d{g}\over dl} = - b_2 g^2 - b_3 g^3, \\
&&{d{u}_k\over dl} =  y_k u_k + c_k g u_k.
\end{eqnarray}
By normalizing appropriately the scaling field $g$, we can also set $|b_2| =
1$. In the case we are considering, $g$ is marginally irrelevant so that $b_2
> 0$ (we assume that $g$ is defined such that $g(l=0) > 0$). We can thus set
$b_2 = 1$. Once this choice has been made, $b_3$ and all coefficients $c_k$
are universal. 

The simple form we have derived above does not strictly apply to the RSIM.
Indeed, in the Ising model the RG operators belong to three different
conformal families and within each family the RG dimensions differ by integers
(see Ref.~\onlinecite{CHPV-02} for a discussion of the irrelevant operators in
the pure Ising model).  Thus, in the present case there are both degeneracies
and resonancies. If we limit our considerations to the relevant scaling
fields, we must only consider the resonance between the identity operator and
the thermal operator, which is responsible for the logarithmic divergence of
the specific heat in the pure Ising model.\cite{Wegner-76} 
By a proper redefinition of the nonlinear scaling fields, one can show that 
in this case the RG equations  (\ref{RGeq-ini}) for the relevant scaling fields 
can be written as:
\begin{eqnarray}
{du_I\over dl} &=& 2 u_I + c_I g u_I + d_I u^2_t, \label{eq-uI}\\
{du_t\over dl} &=& y_t u_t + c_t g u_t, \\
{du_h\over dl} &=& y_h u_h + c_h g u_h, \\
{dg\over dl} &=& - g^2 - b_3 g^3, \label{eq-g}
\end{eqnarray}
where the couplings to the irrelevant scaling fields due to the additional
resonancies have been neglected.  The scaling field $u_I$ is associated with
the identity operator. The additional term $d_I u^2_t$ which appears in 
Eq.~(\ref{eq-uI}) is due to the
resonance with the thermal operator, as in the pure Ising
model.\cite{Wegner-76} The scaling fields $u_t$ and $u_h$ are the relevant
scaling fields associated with the reduced temperature $t$ and the external
field $h$, respectively; $y_t = 1$ and $y_h = 15/8$ are the corresponding RG
dimensions. Finally, $g$ is the marginally irrelevant operator associated with
randomness. The coefficients $c_I$, $c_t$, $c_h$, and $b_3$ are universal,
being independent of the normalization of the scaling fields. By using
conformal field theory, $c_t$, $c_h$, and $b_3$ have been computed:
\cite{Cardy-86,Ludwig-87,LC-87,DPP-95,MK-99}
\begin{equation}
c_t=-1/2, \qquad c_h = 0, \qquad b_3 = -1/2.
\label{ctch}
\end{equation}
Let us now integrate the RG equations. Since $b_3 < 0$, Eq.~(\ref{eq-g})
has two fixed points with $g\ge 0$: one is $g = 0$ and is stable;
the second one is $g = -1/b_3 = 2$ and is unstable. Thus, the basin of 
attraction of the Ising FP corresponds to $g_0=g(l=0) < -1/b_3 = 2$;
for $g_0 > 2$, $g(l)$ flows to infinity. It is important to note that 
Eq.~(\ref{eq-g}) is only valid within the basin of attraction of the 
stable fixed point $g=0$. The redefinitions of the scaling fields 
that we have used to obtain the simple canonical form (\ref{eq-g})
cannot be extended outside the basin of attraction since they are expected 
to become singular at the unstable fixed point. 
The presence of an unstable fixed point indicates that the behavior 
for large values of the disorder is not controlled by the Ising fixed point.
The RG flow could be attracted by a new fixed point---thus, for large values 
of the disorder the transition would be continuous and in a new universality
class---or could go to infinity, indicating the absence of a continuous
transition.  A similar phenomenon was conjectured in three 
dimensions\cite{CPPV-04} on the basis of a perturbative field-theoretical 
analysis of the RG flow.

If $g_0 < -1/b_3$, the function $g(l)$ is 
implicitly given by (we do not replace $b_3$ with its theoretical value
$-1/2$, in order to obtain general expressions that can be tested numerically)
\begin{eqnarray}
&&F[g(l)] = l + F(g_0),\\
&&F(x) \equiv {1\over x} + b_3 \ln \left( x\over 1 + b_3 x\right),
\label{defF}
\nonumber
\end{eqnarray}
The solution can be simplified if we introduce
\begin{equation}
  \tilde{g}(l) = {g(l)\over 1 + b_3 g(l)},
\label{gtilde}
\end{equation}
which satisfies the implicit equation
\begin{eqnarray}
&&\tilde{F}[\tilde{g}(l)] = l + \tilde{F}(\tilde{g}_0), \nonumber \\
&&\tilde{F}(x) \equiv {1\over x} + b_3 \ln x.
\label{defF2}
\end{eqnarray}
This equation can be inverted to give
\begin{equation}
   \tilde{g}(l) = \Phi(\tilde{g}_0,l).
\end{equation}
The function $\Phi(x,l)$ cannot be computed analytically. However, it is
easy to determine it in the large-$l$ limit. We obtain
\begin{equation}
\tilde{g}(l) = {1\over y} - {b_3\ln y\over y^2} + 
         {b_3^2 (\ln^2 y - \ln y)\over y^3} + 
          O\left( {b_3^3 \ln^3 y\over y^4}\right),
\label{gtilde-exp}
\end{equation}
where $y \equiv l + \tilde{F}(\tilde{g}_0)$.
Since $c_h = 0$ the equation for $u_h$ gives 
\begin{equation}
    u_h(l) =  u_{h,0} e^{y_h l},
\end{equation}
where $u_{h,0} = u_h(l=0)$.  In order to determine $u_t(l)$, we rewrite the
corresponding equation as
\begin{equation}
  {d u_t\over u_t} = y_t dl - c_t {g dg\over g^2 + b_3 g^3} = 
         y_t dl - c_t {d\tilde{g}\over \tilde{g}},
\end{equation}
which gives ($c_t = -1/2$, $y_t = 1$)
\begin{equation}
  u_t(l) =  u_{t,0} e^{l} 
   \left[{\tilde{g}(l)\over \tilde{g}_0} \right]^{1/2}\; ,
\label{reweq}
\end{equation}
where $u_{t,0} = u_t(l=0)$.  For large $l$ the function $u_t(l)$ behaves as
\begin{equation}
u_t(l) =  u_{t,0} \tilde{g}_0^{-1/2} {e^l\over l^{1/2}}.
\label{ut-largel}
\end{equation}
Let us finally consider the identity operator. If $d_I = 0$ the solution can
be obtained as in the case of $u_t$. Thus, we write
\begin{equation}
  u_I(l) =  u_{I,0} e^{2l} K(l)
   \left[{\tilde{g}(l)\over \tilde{g}_0 } \right]^{-c_I}\; ,
\end{equation}
where $K(l)$ is an unknown function of $l$, such that $K(l=0) = 1$.
Substituting in the equation for $u_I(l)$ and using the result for $u_t(l)$, we
obtain
\begin{equation}
{dK\over dl} = d_I {u_{t,0}^2\over u_{I,0}} 
        \left[{\tilde{g}(l) \over \tilde{g}_0} \right]^{c_I+1},
\end{equation}
and therefore
\begin{equation}
K(l) = 1 - d_I {u_{t,0}^2\over u_{I,0}} 
         \tilde{g}_0^{-c_I-1}
    \int_{\tilde{g}_0}^{\tilde{g}(l)} dx\, 
         x^{c_I-1} (1 - b_3 x).
\end{equation}
The behavior of $u_I(l)$ for $l\to\infty$ depends on the value of $c_I$. Since
the integral appearing in $K(l)$ diverges as $l^{-c_I}$ for $c_I < 0$, as $\ln
l$ for $c_I = 0$, and is finite for $c_I > 0$, we obtain
\begin{equation}
u_I(l) \sim \cases{ e^{2l} l^{c_I} & $\hphantom{???}$ for $c_I > 0$, \cr
                 e^{2l} \ln l   & $\hphantom{???}$ for $c_I = 0$, \cr
                 e^{2l}  & $\hphantom{???}$ for $c_I < 0$.
    }
\label{uI-largel}
\end{equation}
The RG equations do not fix completely the normalization of the scaling
fields.  First, one can redefine $u_t$, $u_h$, and $u_I$ by a multiplicative
constant;\cite{footnote-norm} such a redefinition is not possible for $g$,
since a multiplicative constant would break the condition $b_2 = 1$. Beside
these trivial redefinitions there is also a nonlinear set of transformations
that leave the equations invariant. Given a constant $\lambda$, we define
$g_\lambda$ as the solution of the equation
\begin{equation}
   F(g_\lambda) = F(g) + \lambda.
\end{equation}
Then, for any $\lambda$ we have 
\begin{equation}
{dg_\lambda\over dl} = - {g_\lambda}^2 - b_3 {g_\lambda}^3.
\end{equation}
Note that the transformation is analytic in a neighborhood of $g = 0$.
If $\tilde{g}_\lambda$ is defined as in Eq.~(\ref{gtilde}), we obtain
\begin{equation}
{\tilde{g}}_\lambda = \tilde{g} [1 - \lambda \tilde{g} + O(\tilde{g}^2)].
\label{gprimo}
\end{equation}
Analogously, if  we define 
\begin{equation}
  u_{t,\lambda} = u_t (\tilde{g}/\tilde{g}_\lambda)^{-1/2},
\end{equation}
then $u_{t,\lambda}$ satisfies the same equation of $u_t$ with $g_\lambda$
replacing $g$, as it can be seen from Eq.~(\ref{reweq}). 
A similar redefinition can be made for $u_I$.
This invariance implies that, beside fixing the normalizations
of $u_t$, $u_h$, and $u_I$, we must also appropriately fix $g$. In practical
terms, $F(g_0)$ is completely arbitrary and must be fixed in order to define
$g(l)$ unambigously. Finally, note that there are no analytic redefinitions of 
$g$ that map  Eq.~(\ref{eq-g}) in an identical equation with 
$b_3' \not= b_3$, proving the universality of $b_3$. 

Neglecting scaling
corrections that are suppressed by power laws, we write the free energy
in the scaling form\cite{Wegner-76}
\begin{equation}
{\cal F}(t,h,p) = e^{-2l} u_I(l) + e^{-2l} f_{\rm sing}(u_t(l), u_h(l), g(l)),
\label{scalft}
\end{equation}
for any $l$.  Note that the whole dependence on $t\equiv T/T_c-1$, $h$, and
$p$ is encoded in the constants $\tilde{g}_0$, $u_{t,0}$, $u_{h,0}$,
$u_{I,0}$, and $d_I$. Of course, $u_{t,0}\sim t$ and $u_{h,0}\sim h$ vanish at
the critical point, while $\tilde{g}_0$ vanishes for $p = 1$.  The
independence of Eq.~(\ref{scalft}) on $l$ allows us to simplify the general
expression for the free energy.  We choose $l$ such that $u_t(l) = 1$ and thus
\begin{eqnarray}
&&e^l = \tau^{-1}
      \left(- \ln \tau \right)^{1/2} 
     \left[1 + O\left({\ln|\ln\tau| \over \ln \tau} \right)\right],
\nonumber \\
&&\tilde{g}(l) = - {1\over \ln \tau} 
     \left[1 + O\left({\ln|\ln\tau| \over \ln \tau} \right)\right],
\end{eqnarray}
where $\tau = u_{t,0}/\tilde{g}_0^{1/2}$.
Substituting these expressions in 
Eq.~(\ref{scalft}), we obtain the general dependence of the 
free energy on $t$ and $h$. 

In order to determine $c_I$, we consider the specific heat $C_h$. The leading
singular behavior is due to the temperature dependence of the scaling field
$u_I$.  Using Eq.~(\ref{uI-largel}) we obtain
\begin{equation}
C_h\propto {\partial^2 {\cal F}(t,0,p)\over \partial t^2}
\sim \cases{ (\ln 1/t)^{c_I} & $\hphantom{???}$ for $c_I > 0$, \cr
               \ln \ln (1/t)   & $\hphantom{???}$ for $c_I = 0$, \cr
               $\rm constant$  & $\hphantom{???}$ for $c_I < 0$.
    }
\label{CH-asyt}
\end{equation}
The asymptotic behavior of the specific heat of 2D randomly diluted Ising 
systems has been determined in Refs.~\onlinecite{DD-81,Shalaev-84,Shankar-87},
obtaining
\begin{equation}
C_h \sim \ln \ln (1/t).
\label{specheatst}
\end{equation}
Comparing with Eq.~(\ref{CH-asyt}), we obtain $c_I = 0$. 
In this case we have
\begin{equation}
u_I(l) = u_{I,0} e^{2l} - {d_I u_{t,0}^2 e^{2l} \over \tilde{g}_0}
   \left[\ln {\tilde{g}(l)\over \tilde{g}_0}  - b_3 (\tilde{g} - \tilde{g}_0)
             \right].
\label{uI-expr}
\end{equation}
It is interesting to note that, since $c_h = c_I = 0$,
randomness couples only to the thermal scaling
field $u_t$.  This result appears quite natural from the point of view of the
Landau-Ginzburg-Wilson approach to critical phenomena. Indeed, in field theory
randomly diluted models are obtained by coupling disorder to the energy
operator:~\cite{Lubensky-75,GL-76}
\begin{equation}
{\cal H} = \int d^d x 
\left\{ {1\over 2}(\partial_\mu \phi(x))^2 + {1\over 2} r \phi(x)^2 
+ {1\over 2} \psi(x) \phi(x)^2  +
{1\over 4!} g_0 \left[ \phi(x)^2\right]^2 \right\},
\label{Hphi4ran}
\end{equation}
where $r\propto T-T_c$, and $\psi(x)$ is a spatially uncorrelated random field
with Gaussian distribution. The 2D RDI critical behavior has been also
investigated by using this field-theoretical approach and the
so-called replica trick. The analysis of the corresponding 
five-loop perturbative expansions\cite{COPS-04,Mayer-89} gives 
results which are substantially
consistent with the marginal irrelevance of disorder.

\subsection{Finite-size scaling}
\label{fss}

Let us now discuss the implications of the above RG analysis for  
the FSS of thermodynamic quantities at the critical point.
We start from the scaling behavior of the free energy
\begin{equation}
{\cal F}(t,h,p,L) = e^{-2l} u_I(l) + e^{-2l} f(u_t(l), u_h(l), g(l), e^lL^{-1}),
\label{scalf}
\end{equation}
where the contributions of the irrelevant scaling fields have been
neglected. By choosing $l=\ln L$, we can write
\begin{equation}
{\cal F}(t,h,p,L) = L^{-2} u_I(\ln L) + L^{-2} f(u_t(\ln L), u_h(\ln L), g(\ln L)).
\label{scalfL}
\end{equation}
If we set $\tilde{F}(\tilde{g}_0) = -\ln L_0$ in
Eq.~(\ref{defF2}),
for $L\to \infty$ we have 
\begin{equation}
\tilde{g}(\ln L) = {1\over \ln L/L_0} \left[
   1 - b_3 {\ln \ln {L/L_0}\over\ln L/L_0}  + 
   O\left({1\over (\ln L/L_0)^2}\right)\right].
\end{equation}
The free energy can then be written as
\begin{eqnarray}
{\cal F}(t,h,p,L) &=& k_1 \ln \tilde{g}(\ln L) + k_2 + 
               k_3 \tilde{g}(\ln L) \nonumber \\
&& + f(u_{t,0} L \tilde{g}(\ln L)^{1/2}, 
       u_{h,0} L^{15/8}, \tilde{g}(\ln L)).
\label{eq-F}
\end{eqnarray}
The constants $k_i$, $u_{t,0}$, $u_{h,0}$, and $L_0$ depend on $t$,
$h$, and $p$. Moreover, $u_{t,0} \sim t$ and $u_{h,0} \sim u_h
\sim h$ close to the critical point. The terms proportional to $k_1$, $k_2$,
and $k_3$ are due to the identity operator, whose dependence on $\tilde{g}(\ln
L)$ is given in Eq.~(\ref{uI-expr}).  Eq.~(\ref{eq-F}) is valid up to
contributions of the irrelevant operators, which are expected to scale as
inverse powers of $L$.

From Eq.~(\ref{eq-F}) we can compute zero-momentum quantities that involve
disorder averages of a single thermal average. For instance, for the specific
heat at $T=T_c$ and $h=0$ we obtain
\begin{equation}
C_h\sim \ln \ln L.
\label{fssch}
\end{equation}
For the susceptibility at $h = 0$ we obtain
\begin{equation}
\chi = \left( {\partial u_{h,0}\over \partial h}\right)^2 L^{7/4}
   f_\chi (u_{t,0} L \tilde{g}(\ln L)^{1/2},\tilde{g}(\ln L)),
\end{equation}
where, as before, we neglect power-law scaling corrections. A similar
scaling equation holds for $U_4$:
\begin{equation}
U_4 = 
   f_{U_4} (u_{t,0} L \tilde{g}(\ln L)^{1/2},\tilde{g}(\ln L)).
\label{scalingU4}
\end{equation}
The determination of the scaling behavior of $U_{22}$ and $R_\xi \equiv \xi/L$
requires an extension of the scaling Ansatz (\ref{eq-F}). A detailed
discussion is presented in Sec. 3.1 of Ref.~\onlinecite{HPPV-07}.  It shows
that both quantities behave as $U_4$, apart from scaling corrections. Thus, if
$R = U_{22}$ or $R = R_\xi$, we also have
\begin{eqnarray}
&& R = 
   f_{R} (u_{t,0} L \tilde{g}(\ln L)^{1/2},\tilde{g}(\ln L)).
\label{scalingRxi}
\end{eqnarray}
Derivatives of the phenomenological couplings have a simple behavior as well,
the leading term being of the form
\begin{equation}
 {\partial R \over d\beta} = 
   \left( {\partial u_{t,0}\over \partial t}\right) L 
      \tilde{g}(\ln L)^{1/2}
   f_{dR} (u_{t,0} L \tilde{g}(\ln L)^{1/2},\tilde{g}(\ln L)).
\end{equation}
At the critical point we can set $u_{t,0} = 0$, so that we can write the 
scaling behaviors
\begin{eqnarray}
&& R =
   g_{R}[\tilde{g}(\ln L)], \nonumber 
\\[2mm]
 && {\partial R \over d\beta} = 
   \left( {\partial u_{t,0}\over \partial t}\right) L 
      \tilde{g}(\ln L)^{1/2}
   g_{dR} [\tilde{g}(\ln L)].
\end{eqnarray}
The functions $g_{R}(x)$ and $g_{dR}(x)$ are universal once an appropriate
normalization is chosen for $\tilde{g}(\ln L)$, which is independent
of the model.  For this purpose, let us
consider a phenomenological coupling $R$. For $L\to\infty$ we can expand
\begin{equation}
    R = R^* + r_1 \tilde{g}(\ln L) + r_2 \tilde{g}(\ln L)^2 + \cdots
\end{equation}
Now we normalize $\tilde{g}(\ln L)$ by requiring $r_2 = 0$.  It is easy to
prove that this is a correct normalization condition. Indeed, imagine that
$\tilde{g}(\ln L)$ has been normalized arbitrarily so that $r_2 \not= 0$.
Then, redefine $\tilde{g}(\ln L)$ by using Eq.~(\ref{gprimo}).  By properly
choosing $\lambda$, it is easy to see that one can set $r_2 = 0$.  
This condition fixes uniquely the scale $L_0$.

Note that, in the pure Ising model, we have $U_{22} = 0$, so that we expect
at the critical point
\begin{equation}
U_{22}\sim \tilde{g}(\ln L)
\label{u22gl}
\end{equation}
for $L\to \infty$. It is natural to invert this relation to express
$\tilde{g}(\ln L)$ in terms of $U_{22}(L)$. Then, we  obtain the scaling forms
\begin{eqnarray}
&& R(L) = \tilde{f}_R(U_{22}) , \label{frl2} \\
&& \chi(L) = d_\chi L^{7/4} \tilde{f}_\chi(U_{22}) , \label{chiU22} \\
&& {\partial R (L)\over d\beta} = d_{dR} L U_{22}^{1/2} \tilde{f}_{dR}(U_{22}),
   \label{dRU22} 
\end{eqnarray}
where $\tilde{f}_R(x)$, $\tilde{f}_\chi(x)$, $\tilde{f}_{dR}(x)$
are universal scaling functions that are normalized such that 
$\tilde{f}_R(0) = R^*$, $\tilde{f}_\chi(0) = \tilde{f}_{dR}(0) = 1$
and have a regular expansion in powers of $x$. Note that these
scaling equations are much simpler than those in terms of $\tilde{g}(\ln L)$,
since they are independent of the scale $L_0$ and of the normalization
of $\tilde{g}(\ln L)$.

\subsection{Finite-size scaling at a fixed phenomenological coupling}
\label{fssfc}

Instead of computing the various quantities at fixed Hamiltonian parameters,
one may study FSS keeping a phenomenological coupling $R$ fixed at a given
value $R_{f}$, as proposed in Ref.~\onlinecite{Hasenbusch-99} and also
discussed in Refs.~\onlinecite{HPV-05,HPPV-07}. This means that, for each $L$,
one considers $\beta_f(L)$ such that
\begin{equation}
R(\beta=\beta_f(L),L) = R_{f}.
\label{rcbeta}
\end{equation}
All interesting thermodynamic quantities are then computed at $\beta =
\beta_f(L)$.  The pseudocritical temperature $\beta_f(L)$ converges to
$\beta_c$ as $L\to \infty$.  

In the next section we report a FSS analysis of MC simulations keeping the
phenomenological coupling $R_\xi$ fixed.  The value $R_{f}$ can be specified
at will as long as it is between the corresponding high-temperature and
low-temperature values.  Since we wish to check the hypothesis that the
asymptotic critical behavior is governed by the Ising fixed point, we choose
$R_{\xi,f}=R_{\rm Is}$, where $R_{\rm Is}=0.9050488292(4)$ is the universal
value of $R_\xi\equiv\xi/L$ at the critical point in the 2D Ising universality
class \cite{SS-00} for square $L\times L$ lattices with periodic boundary
conditions. Note, however, that this choice does not bias our analysis in
favor of the Ising nature of the transition.  By fixing $R_\xi$ to the
critical Ising value, we can perform the following consistency check: if the
transition belongs to the Ising universality class, then any other
RG-invariant quantity must converge to its critical-point value in the Ising
model.

In the $(t,L)$ plane, the line $R_\xi = R_{\rm Is}$ is obtained by solving
the equation
\begin{equation}
 f_{R_\xi} (u_{t,0} L \tilde{g}(\ln L)^{1/2},
               \tilde{g}(\ln L) ) = R_{\rm Is}.
\end{equation}
It gives a relation 
\begin{equation}
    u_{t,0} L \tilde{g}(\ln L)^{1/2} = k(\tilde{g}(\ln L)),
\label{uL-tildeg}
\end{equation}
where $k(x)$ has a regular expansion in powers of $x$. 
Moreover, since we have chosen $R_{\xi,f} = R_{\rm Is}$, 
we have $k(0) = 0$. Substituting relation (\ref{uL-tildeg})
in the above-reported scaling equations for the susceptibility 
$\chi$, the phenomenological couplings $R$, and their derivative, 
we obtain at fixed $R_\xi$
\begin{eqnarray}
&& \chi(L) = c_\chi L^{7/4} C_\chi(\tilde{g}(\ln L)), \\
&& R(L) = C_R(\tilde{g}(\ln L)), 
\label{frl}  \\
&& {\partial R(L)\over d\beta} = 
     c_{dR} L \tilde{g}(\ln L)^{1/2} C_{dR}(\tilde{g}(\ln L)).
\label{chifs}
\end{eqnarray}
The scaling functions are universal, have a regular expansion in powers of
$\tilde{g}(\ln L)$, and are normalized such that $C_R(0)=R_{\rm Is}$,
$C_\chi(0)=C_{dR}(0) = 1$.  The additional corrections due to the irrelevant
operators decay as powers of $1/L$.

The large-$L$ behavior of $\beta_f(L)$ follows from
Eq.~(\ref{uL-tildeg}). Since $k(x) \sim x +  O(x^2)$, we obtain
\begin{equation}
\beta_f - \beta_c = {c_1 \tilde{g}(\ln L)^{1/2} \over L} = 
    {c_1 \over L \sqrt{\ln L/L_0}}
    \left[1 - {b_3\over 2} {\ln\ln L/L_0\over \ln L/L_0} + 
    O\left({1\over \ln L/L_0}\right)\right],
\label{betafap}
\end{equation}
where $L_0$ is computed at the critical point $t=0$.  

We finally mention that Eqs.~(\ref{frl2}), (\ref{chiU22}), (\ref{dRU22}) hold
at fixed $R_\xi=R_{\rm Is}$ as well. The corresponding universal scaling
functions depend on the values of $U_{22}$ at $R_\xi=R_{\rm Is}$ fixed, i.e.,
$\bar{U}_{22}(L) = U_{22}(\beta_f(L),L)$, (we denote them by
$\bar{f}_R(\bar{U}_{22})$, $\bar{f}_\chi(\bar{U}_{22})$, and
$\bar{f}_{dR}(\bar{U}_{22})$, respectively) and have a regular expansion in
powers of $\bar{U}_{22}$.

\section{Finite-size scaling analysis of Monte Carlo simulations}
\label{secMC}

\subsection{Monte Carlo simulations}

\begin{table}
\squeezetable
\caption{
  MC data at fixed $R_\xi=R_{\rm Is}=0.9050488292(4)$.
  For each model and lattice size $L$, we report the number of samples $N_s$, 
  the quartic cumulants $\bar{U}_4$ and $\bar{U}_{22}$,
  the magnetic susceptibility $\chi$,
  the derivative $R'_\xi\equiv dR_\xi/d\beta$, and the specific heat $C_h$.
  If the asymptotic behavior is controlled by the Ising fixed
  point, for $L\rightarrow \infty$ we should have
  $\bar{U}_4\rightarrow U_{{\rm Is}}=1.167923(5)$
  and   $\bar{U}_{22}\rightarrow 0$.
}
\label{tabRxi}
\begin{ruledtabular}
\begin{tabular}{crrllccl}
\multicolumn{1}{c}{Model}&
\multicolumn{1}{c}{$L$}&
\multicolumn{1}{c}{$N_s/10^3$}&
\multicolumn{1}{c}{$\bar{U}_4$}&
\multicolumn{1}{c}{$\bar{U}_{22}$}&
\multicolumn{1}{c}{$\chi$}&
\multicolumn{1}{c}{$R'_\xi$}&
\multicolumn{1}{c}{$C_h$} \\
\colrule
RSIM, $p=0.9$ & 8 & 5361 & 1.16476(3) & 0.05083(3) &  36.1853(9)& 6.5911(11) &  2.7285(5) \\
&16& 2560 & 1.16463(4) & 0.04170(4) & 122.367(4) & 12.608(5) & 3.4282(12)  \\
&32& 1280 & 1.16507(4) & 0.03618(4) & 412.573(14)& 24.029(9) & 4.0283(14) \\
&64& 640 & 1.16563(6) & 0.03237(6)  & 1389.57(8)  & 45.91(3) & 4.550(3) \\
&128& 640 & 1.16597(6) & 0.02947(6) & 4677.0(3) &87.84(7) & 5.014(3) \\ 
&256& 653 & 1.16619(5) & 0.02704(5) & 15741.3(7)  &168.50(9) & 5.431(3) \\
&512& 633 & 1.16656(4) & 0.02522(5) & 52962(2) & 324.18(19) & 5.815(3) \\ 
\colrule
RSIM, $p=0.7$ & 
8& 640 & 1.14557(10) & 0.09561(10) & 25.841(3) &1.2536(13) & 0.2155(3) \\
&16 & 2176 & 1.15206(6) & 0.07526(6) & 86.941(5) &2.6583(11) & 0.30474(14) \\
&32 & 1280 & 1.15682(6) & 0.06297(7) & 293.888(15)&4.967(2) & 0.35203(12) \\
&64 & 658 & 1.15996(7) & 0.05491(8) & 993.26(6)  &9.140(4) & 0.38283(10) \\
&128& 843 & 1.16185(6) & 0.04871(7) &3351.92(14) &16.903(6) & 0.40516(7) \\
&256& 1288& 1.16313(4) & 0.04368(5) & 11299.1(3) &31.501(9) & 0.42262(5) \\ 
\colrule
$\pm J$ Is, $p=0.95$ 
&8 & 3200 & 1.16399(2) & 0.04026(3) & 40.9962(8) & 6.0887(15) & 2.6027(7) \\
&16& 3200 & 1.16405(3) & 0.04023(3) &139.318(5)  &11.163(2) & 3.1527(8)  \\
&32& 3200 & 1.16439(2) & 0.03847(3) &470.511(11) &20.924(3) & 3.6299(5) \\ 
&64& 812 & 1.16482(4) & 0.03592(5) &1585.27(7)  &39.570(10) & 4.0371(7) \\ 
&128& 658 & 1.16527(4)& 0.03331(6) &5337.0(3) &75.12(2) & 4.3865(5) \\
\end{tabular}
\end{ruledtabular}
\end{table}

We perform high-statistics MC simulations of the RSIM at $p=0.9,\,0.7$, and of
the $\pm J$ Ising model at $p=0.95$.  We consider square lattices of linear
size $L$ with periodic boundary conditions.  In the MC simulations of the RSIM
we use a mixture of Metropolis and Wolff cluster\cite{Wolff-89} updates as we
did in the three-dimensional case reported in Ref.~\onlinecite{HPPV-07}. In
the case of the $\pm J$ Ising model, the Wolff cluster update is expected to
be slow\cite{HPPV-07-pmj} so that we only use Metropolis updates with
multispin coding.

Instead of computing the different quantities at fixed Hamiltonian parameters,
we compute them at fixed $R_\xi\equiv \xi/L=R_{\rm Is}$.  This means that,
given a MC sample generated at $\beta = \beta_{\rm run}$, we determine the
value $\beta_{f}$ such that $R_\xi(\beta = \beta_{f}) = R_{f}$.  All
interesting observables are then computed at $\beta = \beta_{f}$.  The
pseudocritical temperature $\beta_{f}$ converges to $\beta_c$ as $L\to
\infty$.  This method has the advantage that it does not require a precise
knowledge of the critical value $\beta_c$ (an estimate is only needed to fix
$\beta_{\rm run}$ that should be close to $\beta_{c}$).  Moreover, for some
observables the statistical errors at fixed $R_\xi$ are smaller than those at
fixed $\beta$ (close to $\beta_{c}$).\cite{HPV-05,HPPV-07} In order to compute
any quantity at $\beta = \beta_f$, we determine its Taylor expansion around
$\beta_{\rm run}$, as we did in our previous work.\cite{HPPV-07-pmj}
Particular care has been taken to avoid any bias due to the finite number of
iterations for each sample: we use the method described in
Ref.~\onlinecite{HPPV-07} and extended to correlated data in
Ref.~\onlinecite{HPPV-07-pmj}.

The results at fixed $R_\xi=R_{\rm Is}$ are reported in
Table~\ref{tabRxi}. For each model and lattice size $L$, we report the number
$N_s$ of samples, the MC estimates of the quartic cumulants $U_4$ and $U_{22}$
at fixed $R_\xi=R_{\rm Is}$ (we denote them with $\bar{U}_4$ and
$\bar{U}_{22}$, respectively), the magnetic susceptibility $\chi$, the
derivative $R_\xi'\equiv dR_\xi/d\beta$, and the specific heat $C_h$.

\subsection{Results} 

\subsubsection{Approach to the 2D Ising fixed-point values}
\label{naivefss}

\begin{figure*}[tb]
\centerline{\psfig{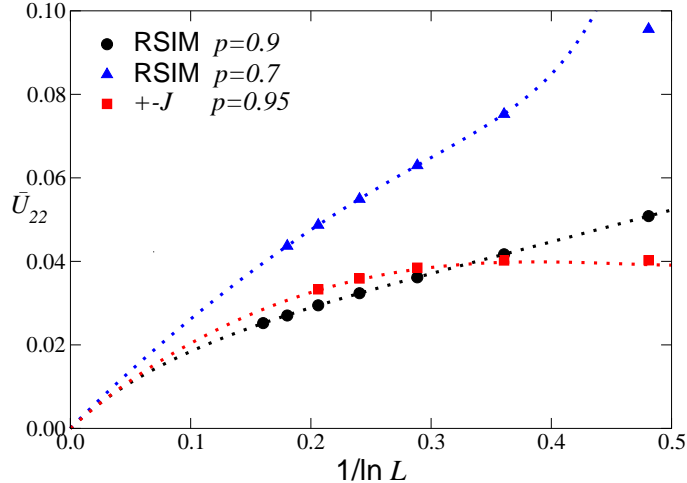}}
\caption{ 
  The phenomenological coupling $\bar{U}_{22}$ vs $1/{\rm ln} L$.  The lines
  show the results of fits to Eq.~(\protect\ref{alogans}).  For the RSIM at
  $p=0.9$ and the $\pm J$ Ising model we fit all data, while for the RSIM at
  $p=0.7$, we use data satisfying $L\ge 16$. Note that the asymptotic  
  slope as $1/\ln L \to 0$ 
  of the resulting curves is identical in the three cases, confirming the
  universality of $C_{22,1}$, defined by $\bar{U}_{22}= C_{22,1} \tilde{g}(\ln
  L) + O(\tilde{g}(\ln L)^3)$, see Sec.~\ref{u22vsL}.  }
\label{fig:u22}
\end{figure*}

Since we perform our FSS analysis keeping $R_\xi=R_{\rm Is}$ fixed, if the
critical behavior is controlled by the Ising fixed point, in the large-$L$
limit we should have
\begin{eqnarray}
\bar{U}_{22}(L) \rightarrow 0, \qquad 
\bar{U}_4(L) \rightarrow U_{\rm Is},
\label{largel}
\end{eqnarray}
where~\cite{SS-00} $U_{\rm Is}=1.167923(5)$ is the universal large-$L$ limit
of the quartic (Binder) cumulant at the critical point in the 2D Ising model.
Since disorder is expected to be marginally irrelevant, see
Sec.~\ref{rgeqrdi}, the approach of $\bar{U}_{22}$ and $\bar{U}_4$ to their
large-$L$ Ising limit is expected to be logarithmic. 

The MC data of $\bar{U}_4$ and $\bar{U}_{22}$, reported in Table~\ref{tabRxi},
clearly approach the Ising values (\ref{largel}). In the case of
$\bar{U}_{4}$, see Table~\ref{tabRxi}, the MC data are very close to $U_{\rm
  Is}=1.167923(5)$. For the largest lattices the relative difference $\delta_4
\equiv |\bar{U}_{4} - U_{\rm Is}|/U_{\rm Is}$ is very small, $\delta_4 \approx
0.0012,\, 0.0041,\,0.0023$ for the RSIM at $p=0.9$ ($L=512$) and $p=0.7$
($L=256$), and the $\pm J$ Ising model at $p=0.95$ ($L=128$), respectively.
However, the asymptotic approach to the large-$L$ Ising value is very slow,
hinting at logarithmic corrections. This is also strongly suggested by the MC
data of $\bar{U}_{22}$, which are shown versus $1/{\rm ln} L$ in
Fig.~\ref{fig:u22}.

\begin{figure*}[tb]
\centerline{\psfig{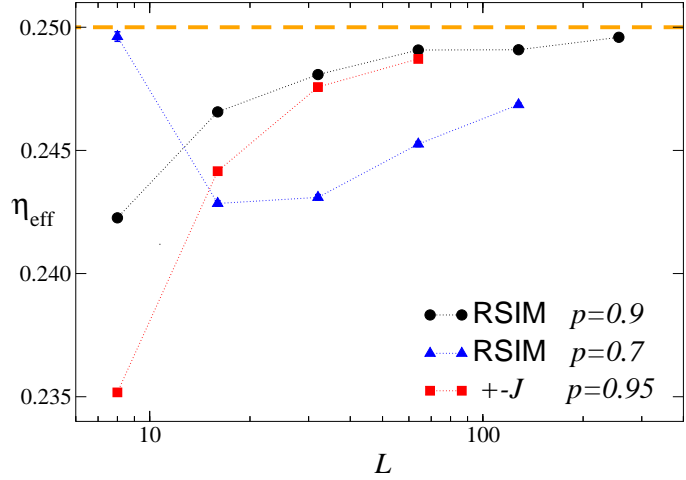}}
\caption{
MC estimates of $\eta_{\rm eff}(L)$.
The dashed line corresponds to the Ising value $\eta=1/4$.
The dotted lines are drawn to guide the eye.
}
\label{effexpeta}
\end{figure*}

\begin{figure*}[tb]
\centerline{\psfig{width=9truecm,angle=0,file=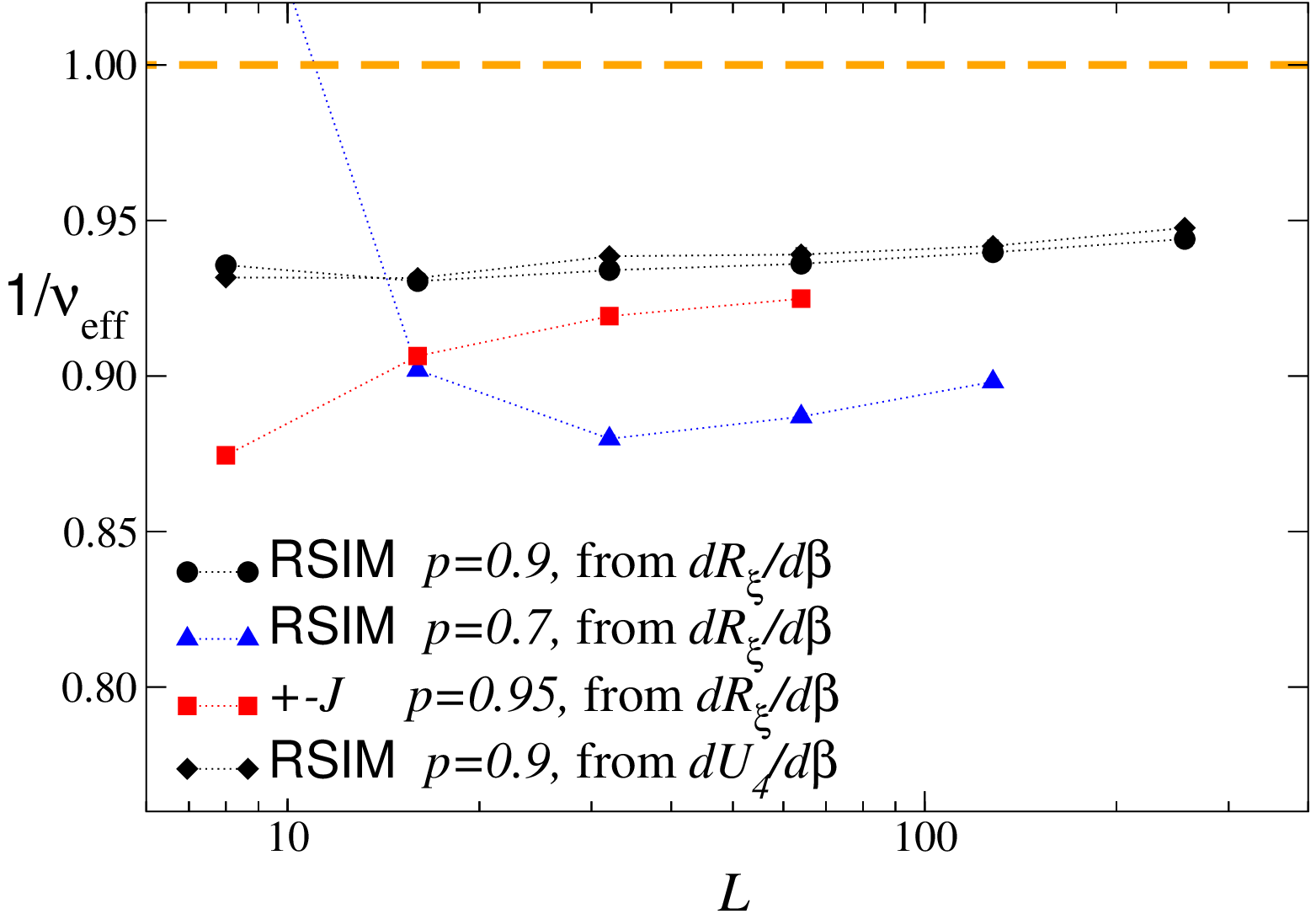}}
\caption{ MC estimates of $1/\nu_{\rm eff}(L)$.  The dashed line corresponds
  to the Ising value $1/\nu=1$.  The dotted lines are drawn to guide the eye.
}
\label{effexpnu}
\end{figure*}

In order to check the approach of the critical exponents
to the Ising values, we define the effective exponents
\begin{equation}
\eta_{\rm eff}(L)\equiv  2 - {\ln \chi(2L)/\chi(L) \over \ln 2},
\label{etaeff}
\end{equation}
and
\begin{eqnarray}
&&1/\nu_{\rm eff}(L) \equiv  {\ln R'_\xi(2L)/ R'_\xi(L)\over \ln 2},
\label{nueffdef}\\
&&1/\nu_{U,\rm eff}(L) \equiv  {\ln U'_4(2L)/U'_4(L)\over \ln 2},
\label{nueffdef2}
\end{eqnarray}
where we indicate the derivative with respect to $\beta$ with a prime.  The MC
estimates of $\eta_{\rm eff}(L)$ and $1/\nu_{\rm eff}(L)$ are plotted in
Figs.~\ref{effexpeta} and \ref{effexpnu}.  They appear to approach the Ising
values $\eta=1/4$ and $1/\nu=1$.  In the case of $\eta$, the raw data are
already very close to the Ising value: the largest lattices give $\eta_{\rm
  eff}(L=256)=0.24959(8)$ for the RSIM at $p=0.9$, $\eta_{\rm
  eff}(L=128)=0.24686(8)$ for the RSIM at $p=0.7$, and $\eta_{\rm
  eff}(L=64)=0.24871(10)$ for the $\pm J$ Ising model at $p=0.95$.  In the
case of $1/\nu_{\rm eff}(L)$ the approach is much slower.  At the largest
available lattices we find $1/\nu_{\rm eff}(L=256)=0.9441(12)$ for the RSIM at
$p=0.9$, $1/\nu_{\rm eff}(L=128)=0.8981(7)$ for the RSIM at $p=0.7$, and
$1/\nu_{\rm eff}(L=64)=0.9249(5)$ for the $\pm J$ Ising model at $p=0.95$.
Anyway, all data show an upward trend towards the Ising value $1/\nu=1$.

These results provide already a quite strong evidence that the asymptotic
behavior of the FSS is universal and it is controlled by the Ising fixed
point, with scaling corrections which decay very slowly.  In the following we
report a more careful analysis of these logarithmic corrections, showing that 
they have a universal pattern which is consistent with the RG predictions
obtained in Sec.~\ref{RGFSS}.

\subsubsection{Universal finite-size behavior as a function
of the phenomenological coupling $U_{22}$}
\label{u22fss}

As discussed in Sec.~\ref{fss}, the FSS formulas obtained from the RG
equations of Sec.~\ref{rgeqrdi} can be written in terms of the
phenomenological coupling $U_{22}$.
In the following we compare the MC
data with the predictions reported in Sec.~\ref{fss} and \ref{fssfc}, and in
particular with Eqs.~(\ref{frl2}), (\ref{chiU22}), and (\ref{dRU22}).

\begin{figure*}[tb]
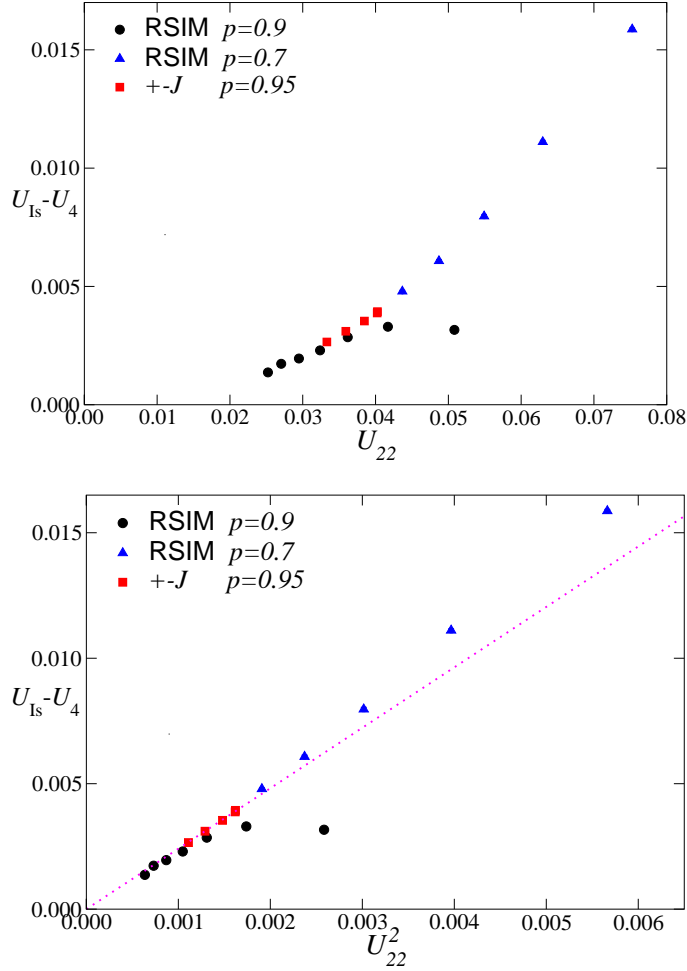

\centerline{\psfig{width=9truecm,angle=0,file=u4u22.eps}}
\vspace{4mm}
\centerline{\psfig{width=9truecm,angle=0,file=u4u22b.eps}}
\caption{
$U_{\rm Is}-\bar{U}_4$ vs $\bar{U}_{22}$ (above)
and $\bar{U}_{22}^2$ (below). 
}
\label{u4u22}
\end{figure*}

Let us first consider the quartic cumulant $\bar{U}_4$ defined in
Eq.~(\ref{cumulants}).  At fixed $R_\xi$, $\bar{U}_4(L)$ is expected to behave
as
\begin{equation}
\bar{U}_4(L) = \bar{f}_{U_4}(\bar{U}_{22}),
\label{beh-barU4}
\end{equation}
where $\bar{f}_{U_4}(x)$ is a universal function, analytic at $x=0$,
satisfying $\bar{f}_{U_4}(0) = U_{{\rm Is}}$.  Corrections to the behavior
(\ref{beh-barU4}) vanish as powers of $1/L$.  The scaling behavior
(\ref{beh-barU4}) is well satisfied by the MC data, as shown in
Fig.~\ref{u4u22}. All data fall on a single curve, except for a few of them
corresponding to small values of $L$ (this is particularly evident in the data
for the RSIM at $p=0.9$), indicating the presence of power-law scaling 
corrections. The results show that the linear term is absent or negligible 
in the expansion of $\bar{f}_{U_4}(\bar{U}_{22})$ around 
$\bar{U}_{22} = 0$; if the data are plotted versus 
$\bar{U}_{22}^2$, they fall on an approximately straight line, 
suggesting that
\begin{equation}
\bar{U}_4(L) - U_{\rm Is} = c \,\bar{U}_{22}(L)^2 + O(\bar{U}_{22}^3).
\label{asu4}
\end{equation}
A fit of the numerical results to 
$\bar{U}_4(L) - U_{\rm Is}=c \,\bar{U}_{22}(L)^ 2$
gives $c = 2.4(2)$. This implies that
\begin{equation}
\bar{U}_{4} (L) = U_{{\rm Is}} + {c_{4}\over (\ln L/L_0)^2} + \ldots
\label{scu4}
\end{equation}
where $L_0$ is the model-dependent constant that appears in the expansion of
$\tilde{g}(\ln L)$ (as such, it is independent of the quantity that one is
considering).

\begin{figure*}[tb]
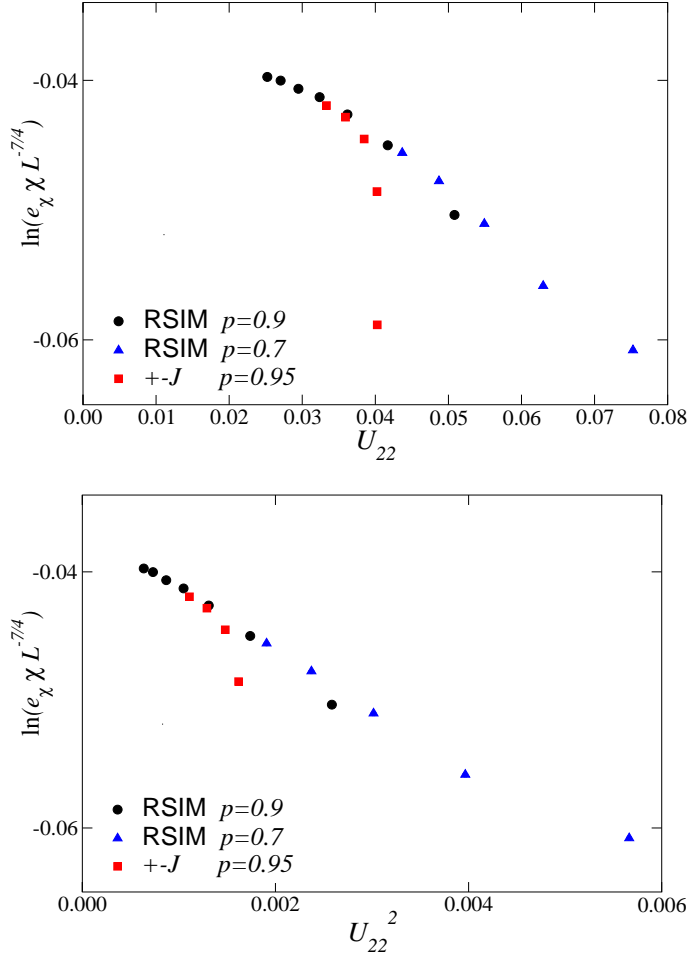

\centerline{\psfig{width=9truecm,angle=0,file=chiu22.eps}}
\vspace{4mm}
\centerline{\psfig{width=9truecm,angle=0,file=chiu22sq.eps}}
\caption{
Plot of $\ln (e_\chi \chi L^{-7/4})$ vs $\bar{U}_{22}$ (top) and 
vs $\bar{U}_{22}^2$ (bottom); we set
$e_\chi = 1,1.4,0.88$ for the RSIM at $p=0.9$ and $p=0.7$, and for 
the $\pm J$ Ising model. 
The constants $e_\chi$ have been chosen
such as to obtain the best collapse of the MC data.
}
\label{chivsU22}
\end{figure*}

As discussed in Secs.~\ref{fss} and \ref{fssfc}, at fixed $R_\xi$, $\chi$
behaves as
\begin{equation}
\chi = d_\chi L^{7/4} \bar{f}_\chi(\bar{U}_{22}(L)),
\label{scalchiu22}
\end{equation}
where $\bar{f}_\chi(x)$ is a universal function such that $\bar{f}_\chi(0) =
1$.  This means that, by properly choosing constants $e_\chi$, the combination
$e_\chi \chi L^{-7/4}$ is a universal function of $\bar{U}_{22}$. In
Fig.~\ref{chivsU22} we show this quantity. The plot is clearly consistent with
Eq.~(\ref{scalchiu22}).  Note also that if the data are plotted versus
$\bar{U}_{22}^2$ they approximately fall on a straight line, suggesting
$\bar{f}_\chi(x)= 1 + O(x^2)$, analogously to the case of $\bar{U}_4$.

\begin{figure*}[tb]
\centerline{\psfig{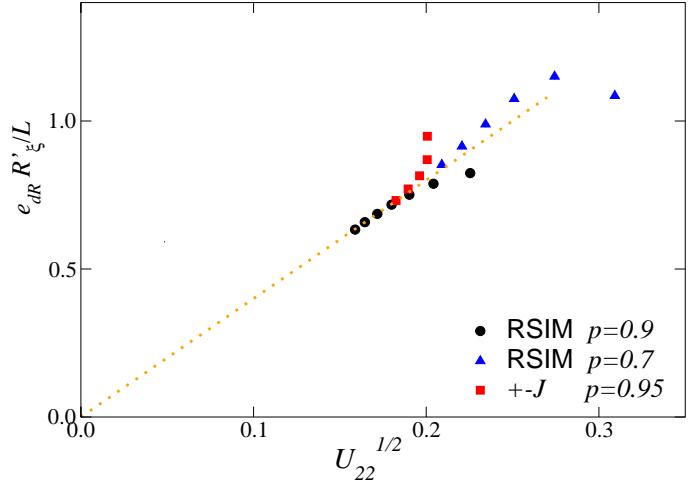}}
\caption{
  Plot of $e_{dR} R'_\xi/L$ versus $\bar{U}_{22}^{1/2}$. We have chosen
  $e_{dR} = 1,6.9,1.2$ for the RSIM at $p=0.9$ and $p=0.7$, and for the $\pm
  J$ Ising model.  The constants $e_{dR}$ have been chosen such as to obtain
  the best collapse of the MC data.  
}
\label{RpvsU22}
\end{figure*}

In Fig.~\ref{effexpnu} we showed the effective exponents (\ref{nueffdef}) and
(\ref{nueffdef2}) related to the thermal exponent $\nu$.  The data approached
the Ising value $\nu_{\rm Is}=1$ with slowly decaying corrections.
The effective exponents computed by using 
Eqs.~(\ref{nueffdef}) and (\ref{nueffdef2}) were
very close, as shown in Fig.~\ref{effexpnu} for the RSIM at $p=0.9$ (this is
also true for the other model considered).  For this reason, 
in the following we focus on $R_\xi'$.  As
discussed in Sec.~\ref{fss} and ~\ref{fssfc}, 
the derivative $R_\xi'$ at fixed $R_\xi$ scales as 
\begin{equation}
R_\xi'= d_{dR} L \bar{U}_{22}(L)^{1/2} \bar{f}_{dR}(\bar{U}_{22}(L)),
\label{scalRpu22}
\end{equation}
where $\bar{f}_{dR}(x)$ is a universal function.  This means that, by properly
choosing some constants $e_{dR}$, the combination $e_{dR} R_\xi'/L$ is a
universal function of $\bar{U}_{22}$. In Fig.~\ref{RpvsU22} we show such
quantity. The plot is clearly consistent with Eq.~(\ref{scalRpu22}): the data
fall on a single curve and approach zero as $\bar{U}_{22}(L)^{1/2}$ when
$\bar{U}_{22}\to 0$.  Again, note the presence of power-law corrections for
large values of $\bar{U}_{22}$.

The approach of $\beta_f(L)$ to $\beta_c$ is given by Eq.~(\ref{betafap}).
Equivalently, we can also consider the scaling form
\begin{equation}
\beta_f(L) - \beta_c = {d_1 \bar{U}_{22}(L)^{1/2}\over L} + 
   {d_2 \bar{U}_{22}(L)^{3/2}\over L} + \cdots
\label{betafc2}
\end{equation}
We determine $\beta_c$ by performing fits to Eq.~(\ref{betafc2}).  We include
only data such that $L\ge L_{\rm min}$, where $L_{\rm min}$ is the smallest
cutoff which provides fits with $\chi^2/{\rm DOF} \lesssim 1$.  Moreover, as a
check we have also performed fits to Eq.~(\ref{betafc2}) in which we only
consider the leading term (i.e. we set $d_2 = 0$). For the RSIM at $p =0.9$ we
obtain $\beta_c = 0.525838(1)$ (fit with $d_2=0$) and $\beta_c = 0.525835(2)$,
if both terms are taken into account. Analogously, these two fits give
$\beta_c = 0.93294(1)$, 0.93289(3) for the RSIM at $p=0.7$ and $\beta_c =
0.53362(1)$, 0.53348(2) for the $\pm J$ Ising model.  Our final estimates are
$\beta_c=0.525835(3),\,0.93289(5),\,0.5335(1)$ respectively for the RSIM at
$p=0.9$ and $p=0.7$, and the $\pm J$ Ising model at $p=0.95$.  Consistent
results are obtained by fitting the data of $\beta_f(L)$ to
Eq.~(\ref{betafap}).

\subsubsection{Universal logarithmic corrections as a function of $L$}
\label{u22vsL}

In the following we directly check the dependence on $L$ of
$\bar{U}_{22}$, $R'_\xi$, and of the specific heat $C_h$.
As discussed in Sec.~\ref{fssfc}, for $L\to \infty$ the phenomenological
coupling $\bar{U}_{22}$ behaves as
\begin{equation}
\bar{U}_{22}(L) = C_{22,1} \tilde{g}(\ln L) + O(\tilde{g}^3),
\end{equation}
where $C_{22,1}$ is a universal constant. The absence of the term of 
order $\tilde{g}^2$ fixes uniquely the normalization of the coupling 
$\tilde{g}$. This quantity can be expanded in powers of $1/\ln (L/L_0)$ to 
different orders. 
Using the expansion (\ref{gtilde-exp}) with $y = \ln L/L_0$,
we can perform three different types of fit, corresponding to
three different approximations for $\tilde{g}(\ln L)$. In fit (a), we fit 
$\bar{U}_{22}(L)$ to 
\begin{equation}
\bar{U}_{22}(L) = {C_{22,1}\over \ln L/L_0},
\end{equation}
where $C_{22,1}$ and $L_0$ are free parameters. In fit (b), we also include 
the next term proportional to $b_3$, i.e. we fit the data to 
\begin{equation}
\bar{U}_{22}(L) = {C_{22,1}\over \ln L/L_0} - 
         {C_{22,1} b_3 \ln \ln L/L_0 \over (\ln L/L_0)^2} ,
\label{alogans}
\end{equation}
where $C_{22,1}$, $b_3$, and $L_0$ are free parameters.
Finally, we can also include the 
next term obtaining [fit (c)]
\begin{equation}
\bar{U}_{22}(L) = C_{22,1}\left\{ {1\over \ln L/L_0} - 
         {b_3 \ln \ln L/L_0 \over (\ln L/L_0)^2}  + 
         {b_3^2[ (\ln \ln L/L_0)^2 - \ln \ln L/L_0] \over (\ln L/L_0)^3}
         \right\},
\end{equation}
where $C_{22,1}$, $b_3$, and $L_0$ are free parameters.  The results of the
fits for different values of $L_{\rm min}$ are reported in Table
\ref{tabfitU22}.  Let us consider first the fit of the data for the RSIM at
$p=0.9$ for which we have the largest lattices. Fit (a) has a very poor
$\chi^2$, indicating that the data are not well fitted by a single logarithmic
term.  If we include the next correction the $\chi^2$ drops dramatically,
indicating that our results are precise enough to be sensitive to the elusive
$\ln \ln L/L_0$ terms. Beside the very good $\chi^2$, the results are also
very stable with $L_{\rm min}$.  This stability should not be trusted too much
however, since fit (c)---which {\em a priori} should better since we include
an additional set of corrections---has a very poor $\chi^2$ and gives results
that vary somewhat with $L_{\rm min}$. As an additional check we also fit
$\bar{U}_{22}(L)$ to
\begin{equation}
  \bar{U}_{22}(L) = C_{22,1} \tilde{g}(\ln L) + 
            C_{22,3} \tilde{g}(\ln L)^3,
\label{fit-c}
\end{equation}
using the expansion of $\tilde{g}(\ln L)$ used in fit (c). For $L_{\rm min} =
8,16,32$ we obtain $\chi^2$/DOF = 213/3, 135/2, 16/1; they are better than
those obtained in fit (c), but still significantly worse than those obtained
in fit (b). Correspondingly, we obtain $C_{22,1} = 0.210(1), 0.254(3),
0.268(10)$ and $b_3 = 1.44(1), 0.89(10), 1.0(3)$ for the same values of
$L_{\rm min}$.  Finally, we fit $\bar{U}_{22}(L)$ to Eq.~(\ref{fit-c}) by
using the exact expression for $\tilde{g}(\ln L)$: for each $L/L_0$,
$\tilde{g}(\ln L)$ is obtained by inverting $\tilde{F}(\tilde{g}) = \ln
L/L_0$, where $\tilde{F}(x)$ is defined in Eq.~(\ref{defF2}).  If we only
include the leading term, i.e.~we set $C_{22,3} = 0$, the quality of the fit
is significantly worse than that of fit (b) and better than that of fit (c):
$\chi^2/{\rm DOF} = 515/4,52/3,6/2$ for $L_{\rm min} = 8,16,32$.
Correspondingly, $C_{22,1} \approx 0.27, 0.29, 0.31$ and $b_3 \approx 2.0,
2.4, 2.7$.  Though the scatter of the estimates of $C_{22,1}$ is significantly
larger than the statistical errors---this should be expected since
$\chi^2$/DOF is significantly larger than 1 in most of the cases---the data
show a clear pattern. If we take the central estimate from fit (b), we obtain
$C_{22,1} \approx 0.28$.  To estimate a reliable error, note that all results
of the fits with $L_{\rm min} \ge 16$ lie in the interval $0.23 \lesssim
C_{22,1} \lesssim 0.31$. A conservative error is therefore $\pm 0.05$, so that
$C_{22,1} = 0.28(5)$. It is more difficult to estimate $b_3$, since this
parameter varies significantly from one fit to the other.  In any case, note
that all results satisfy $b_3 > 0$, in contrast with the theoretical
prediction $b_3 = -1/2$. It is not clear if this difference should be taken
seriously. It might be that it is only an indication that we are not
sufficiently asymptotic to estimate correctly the coefficient of the 
slowly varying $\ln\ln L/L_0$ term.

Since $C_{22,1}$ is universal, we can check its estimate by comparing the
above-reported results with those obtained in the other two models, for which
we have less data.  For both models, fit (a) is significantly worse than fit
(b) or fit (c). For the RSIM at $p=0.7$ fit (b) and fit (c) have similar
reliability. The corresponding estimates of $C_{22,1}$ are fully consistent
with that reported above.  For the $\pm J$ Ising model, only fit (c) is
reliable. The estimates of $C_{22,1}$ are again consistent with those obtained
in the RSIM. The universality of the leading logarithmic correction is well
satisfied by our data.

The scale $L_0$ is very poorly determined and varies significantly with
$L_{\rm min}$ and the type of fit.  The ratio of the scales can also be
determined by directly matching the numerical data. If power-law scaling
corrections are negligible, we should have
\begin{equation}
\bar{U}_{22,{\rm model\,1}}(L) = \bar{U}_{22,{\rm model \,2}}(\kappa L)
\label{matching}
\end{equation}
for some constant $\kappa$, which is the ratio of the scales $L_0$ pertaining
the two models. By direct comparison we obtain $L_{0,{\rm RSIM},p=0.7} \approx
\kappa L_{0,{\rm RSIM},p=0.9}$, $\kappa\gtrsim 16$, and $L_{0,\pm J} \approx
\kappa L_{0,{\rm RSIM},p=0.9}$, with $2 \lesssim \kappa \lesssim 4$.  Since
$L_0$ is independent of the observable, these relations should not be specific
of $\bar{U}_{22}$ but should apply to any RG invariant quantity: indeed, as
can be seen from the data reported in Table~\ref{tabRxi}, they also
approximately hold for $\bar{U}_4$.  Note that $L_0$ increases with $p$ in the
RSIM as expected: the Ising critical behavior is observed for $L\gtrsim L_{\rm
  min}$, with $L_{\rm min}$ increasing with $p$.

\begin{table}
\squeezetable
\caption{Results of the fits. We do not report the results of 
fit (b) for the $\pm J$ Ising model with $L_{\rm min} = 16$ because 
this fit is unstable (apparently, the $\chi^2$ continuously decreases as 
$b_3\to -\infty$ and $L_0\to 0$). DOF is the number of degrees of 
freedom of each fit.
}
\label{tabfitU22}
\begin{ruledtabular}
\begin{tabular}{ccccccccc}
& \multicolumn{2}{c}{Fit (a)}
& \multicolumn{3}{c}{Fit (b)}
& \multicolumn{3}{c}{Fit (c)} \\
\multicolumn{1}{c}{$L_{\rm min}$}&
\multicolumn{1}{c}{$\chi^2$/DOF}&
\multicolumn{1}{c}{$C_{22,1}$}&
\multicolumn{1}{c}{$\chi^2$/DOF}&
\multicolumn{1}{c}{$C_{22,1}$}&
\multicolumn{1}{c}{$b_3$}&
\multicolumn{1}{c}{$\chi^2$/DOF}&
\multicolumn{1}{c}{$C_{22,1}$}&
\multicolumn{1}{c}{$b_3$}
\\
\colrule
& \multicolumn{8}{c}{RSIM $p = 0.9$} \\
8 & 1844/5 & 0.193(1) &  3.4/4 & 0.280(2) & 1.35(1) & 1280/4 & 0.222(1) &0.91(3)
\\
16 & 221/4 & 0.227(1)  &  3.1/3 & 0.281(3) & 1.36(3) & 164/3 & 0.240(2) &0.85(7)
\\
32 & 27/3 & 0.235(2)  &  3.1/2 & 0.281(5) & 1.36(8) & 20/2 & 0.252(5) &0.88(23)
\\
\colrule
& \multicolumn{8}{c}{RSIM $p = 0.7$} \\
8 & 748/4 & 0.276(1) &  15/3 & 0.356(2) & 0.88(2) & 37/3 & 0.334(1) &1.09(1)
\\
16 & 95/3 & 0.287(1)  & 14/2 & 0.350(5) & 0.83(5) & 1.7/2& 0.324(1) &1.30(1)
\\
32 & 0.6/2 & 0.297(1)  &0.3/1 & 0.28(3) & $-$0.3(7)& 0.3/1 & 0.28(3) &$-$0.3(5)
\\
\colrule
& \multicolumn{8}{c}{$\pm J$ model} \\
8 & 4211/3 & 0.986(4) & 2753/2 & 0.610(4) & 2.01(1) & 27/2 & 0.345(3) &1.90(2)
\\
16 &389/2 & 0.449(4)  & -- & -- & -- & 0.02/1& 0.315(6) &1.79(2)
\\
\end{tabular}
\end{ruledtabular}
\end{table}

In order to check the $L$-dependence of the derivative
$R_\xi'$, previosly discussed as a function of $\bar{U}_{22}$, we perform
fits of the MC data of $R_\xi'$ to the behavior
\begin{equation}
\ln {R_\xi'\over L} = a_1 \ln\ln {L\over L_0} + a_2 + 
           {a_3 \ln\ln {L/L_0}\over \ln {L/L_0}} + {a_4 \over \ln {L/L_0}},
\end{equation}
taking $a_1,\ldots,a_4$, and $L_0$ as free parameters.  According to theory,
we should find $a_1 = -1/2$. Because of the presence of 5 free parameters this
fitting form can be safely used only for the RSIM at $p = 0.9$. If we fit all
data we obtain $a_1 = -0.53(5)$ ($\chi^2/{\rm DOF} = 0.71$); if we discard the
result corresponding to $L = 8$ we obtain $a_1 = -0.44(11)$. These results are
in good agreement with the theoretical prediction $a_1 = -1/2$.

\begin{figure*}[tb]
\centerline{\psfig{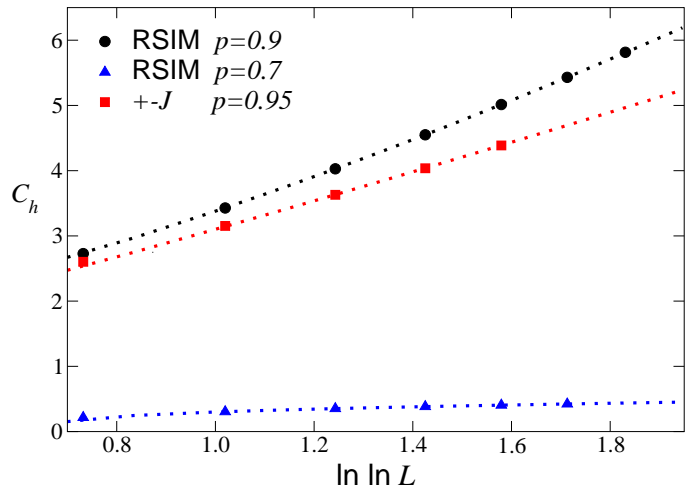}}
\caption{
MC data of the specific heat versus $\ln\ln L$.
The lines correspond to fits to Eq.~(\ref{chsc}).
}
\label{cv}
\end{figure*}

Finally, we consider the specific heat $C_h$,
\begin{equation}
C_h = {[\langle {\cal H}^2 \rangle - \langle {\cal H} \rangle^2]\over V},
\end{equation}
where ${\cal H}$ is the Hamiltonian.
The RG analyses of Refs.~\onlinecite{DD-81,Shalaev-84,Shankar-87,MK-99}
predict a diverging $\ln \ln L$ asymptotic behavior.  In Fig.~\ref{cv} we show
the MC data of $C_h$ at fixed $R_\xi = R_{\rm Is}$. 
They are definitely consistent with the theoretical
prediction for its asymptotic behavior
\begin{equation}
C_h \approx A \ln \ln (L/L_0) + B,
\label{chsc}
\end{equation}
where $A$ and $B$ are nonuniversal parameters, 
and $L_0$ is the model-dependent scale.
Fits of the data to Eq.~(\ref{chsc}), taking $A$, $B$, and $L_0$ 
as free parameters, suggest $A\approx 5, \, 0.1,\, 2$
for the RSIM at $p=0.9$ and $p=0.7$, and the $\pm J$ Ising model
at $p=0.95$, respectively. 
There is a large difference between the values for RSIM at
$p=0.9$ and $p=0.7$, but this should not be surprising because the $p=0.7$ is
quite close to the percolation point~\cite{NZ-00} $p=p_{\rm perc}\approx
0.59$.

%In conclusion, the FSS analysis presented in this section provides a robust
%evidence of the marginal irrelevance of the quenched disorder in the 2D RSIM
%and the $\pm J$ Ising model (for sufficiently small values of $1-p$).  The FSS
%behavior is controlled by the Ising fixed point, with logarithmic corrections.
%Such logarithmic corrections appear universal, and in agreement with the
%results derived from the RG equations (\ref{gop}).

\acknowledgments

We thank Pasquale Calabrese for very useful discussions.


\begin{thebibliography}{99}
\bibitem{footnote-norm}
The normalization of the scaling fields $u_t$, $u_h$, and 
$u_I$ can be fixed by setting,
e.g., $u_{t,0} \approx t$, $u_h\approx h$ close at the critical point, 
and $d_I = 1$. We do not make this choice, since it does not 
simplify the calculations that follow.
\end{thebibliography}
\end{document}